# New Insights into the Dynamics of Swarming Bacteria: A Theoretical Study

David Hansmann[1,2*], Guido Fier[2], Rubén C. Buceta[1,2]

[1] *Departamento de Física, Universidad Nacional de Mar del Plata, Funes 3350, B7602AYL Mar del Plata, Buenos Aires, Argentina*

[2] *Instituto de Investigaciones Físicas de Mar del Plata (IFIMAR), Funes 3350, B7602AYL Mar del Plata, Buenos Aires, Argentina*

*E-mail address: David.Hansmann@conicet.gov.ar

In the present work we simulate the basic two-dimensional dynamics of swarming *E. coli* bacteria on the surface of a moderately soft agar plate. Individual bacteria are modelled by self-propelled ridged bodies (agents), which interact with each other only through inelastic collision and with the highly viscous environment through damping forces. The motion of single agents is modelled closely corresponding to the behaviour of swimming bacteria. The dynamics of the model were adjusted to reproduce the experimental measurements of swimming *E. coli* K-12. Accordingly, simulations with loosely packed agents ($\rho \approx 0$) show typical run-and-tumble statistics. In contrast, simulations with densely packed agents ($\rho \approx 0.3$-$0.7$) are dominated by interactions (collisions) between agents which lead to the emergence of swarming behaviour. In addition, we model the motion of single agents on the base of modified run-and-tumble dynamics, where the bacteria do not turn actively during the tumble. We show that simulations with densely packed modified agents lead as well the emergence of swarming behaviour, if rotational diffusion is considered.

## I. INTRODUCTION

During the last years the physical aspects of individual and collective motion of bacteria attracted an increasing attention of the scientific community[1]. But already in the early 1970s first studies concerning bacterial movements involving statistical physics were performed[2,3]. The complex movement patterns which have been found were analysed and classified by Henrichsen in 1972[4]. According to Henrichsens studies[4], flagella driven bacterial movements can be classified to swimming and swarming movements. Both classes, swarming and swimming, use the same mechanism which is based on a rotating helical flagella generating propulsion thrust[5,6]. The most apparent difference between swimming and swarming is the packing density of bacteria (surface coating). Swimming is an individual movement which takes place at such low population densities that collisions among bacteria can be neglected. On the contrary, swarming is a collective movement which takes place at rather high population densities where collisions among bacteria are omnipresent[4]. Compared to the swimming motion, which has been studied in volumetric liquids (3D) and on porous surfaces (quasi 3D)[7–9], the swarming movement only has been studied in thin, nutrient rich, liquid layers which are formed on surfaces or between two closely-opposed surfaces, e.g. moderately soft agar plates[10]. In addition, swimming bacteria present a characteristic pattern of movement called "Run and Tumble". Here, a bacterium moves in two consecutive modes: The "run" mode where it moves in a good approximation to a straight line and the "tumble" mode where it stops abruptly, turns and changes its direction before continuing to "run". Though the averaged movement direction of swimming bacteria is typically not randomized, but influenced by chemotaxis which allows the bacteria to follow the concentration gradient of attractants or to move away from repellents. More recent studies suggest that the capability of chemotaxis is as well related to the detection and response to extra-cellular signalling molecules (auto-inducers)[11]. This detection and response to auto-inducers is typically referred to as quorum-sensing and is one possible mechanism that regulates the genetic expression of bacteria depending on their population density. In this context is quorum sensing a mechanism which is attributed to the changes that bacteria undergo when bacteria prepare for swarming[12]. These cellular changes are typically hyper-flagellation, elongation of the cell and production of surfactant[10]. It turns out, that due to these changes swarming bacteria colonies enhance spreading which allows bacteria to invade the host tissues and colonize surfaces much faster[10].

In fact, many different bacterial species exhibit both swimming and swarming behaviour depending strongly on the medium the bacteria are inoculated on (e.g. softness of the Agar medium). A prominent representative of those species is *E. coli* which plays an important role in the daily life of humans. In this context, it should be noted that the enhanced bacterial spreading and the high population densities of swarming colonies lead to an increased virulence and greater adaptability to antibiotics[13]. Simply on account of bacterial collisions, it is hardly surprising that "run and tumble" motions are not observed in such fast and powerful moving, dense packed swarming colonies[14]. Additionally, the absence of "run and tumble" could be the result of a genetic suppression in terms of "switching off" run and tumble. Nevertheless the fact that "run and tumble" motion cannot be observed does not mean that it is "switched off". It could rather be that the "run and tumble" movements are disrupted and superimposed as a result of interactions (collisions) between the bacteria in the densely packed colony.

The aim of this study is to address the presence of "run and tumble" in swarming bacterial colonies. We prepared/created and implemented a simple mechanical simulation of a densely packed bacterial colony to analyse its dynamics by means of statistical methods. As stated by other researchers, in general terms, the interaction between the individual agents may play a decisive role in such a simulation[15,16]. However, the interactions in bacterial swarms are rather "simple" compared to interactions that take place in highly developed species. In particular we assumed that a single bacterium does not actively analyse its neighbourhood in order to synchronize its movement with the swarm, but it follows a given motion pattern independent from the surrounding bacteria. This movement pattern may, however, be passively altered by short range interactions like short range hydrodynamic forces or direct collisions among bacteria. As Darton et al. observed[14], independently on that, interactions are restricted to immediate neighbours and can therefore be modelled as direct collision among bacteria[14]. Therefore, using all these assumptions, we examined the impact of the bacterial packing density on their movement patterns, in particular on run and tumble in order to explore insights of these interesting systems.

In the present work we employed self-propelled agents interacting with each other only via collision on a 2-dimensional surface. The interacting potentials were kept as simple as possible and were modelled like not overlapping, not deformable volumes. All agents had an attributed mass, shape (contour), position, velocity, moment of inertia, orientation angle, and angular velocity. Using the later parameter we modelled the interactions like inelastic collisions of non-deformable massive shapes. In addition we assumed very strong dumping forces to simulate the low Reynolds number of the environment, in agreement with Ref [17]. Therefore, all movements in the present systems, i.e., self-propelled agents, were based on simple rigid-body dynamics in a highly viscous medium. As a first step, the simulation was tuned to reproduce the individual movements (run and tumble) of swimming *E. coli* for very low population density using the experimental data of Berg and Brown[3] as a reference. Then, on that basis, the population density was increased and compared to the results the experimental data of

Darton[14]. In addition, the entire work was repeated using modified run and tumble movements, where the active rotations of the bacteria during the tumble were turned off.

## II. METHODS

In this work we studied the dynamics of swarming bacteria on the basis of ellipse-shaped, self-propelled ridged bodies (agents), which interact with each other via inelastic collisions. The dynamics of an individual, freely moving agent was tuned in order to reproduce the experimentally measured run and tumble movements of swimming *E. coli* K-12[3]. The agents were placed on a 2-dimensional plane with periodic boundary conditions simulating a moderately soft agar medium.

The length distribution of the agent population was chosen according to the literature values for swimming and swarming *E. coli*, respectively[14]. Here, the parameter "length" was interpreted as the extension of the agent along its major axis. The mass and the mass distribution inside of the agents were arbitrary quantities and were used as free model parameters. Nevertheless both were crucial for the success of the simulation since they influence the acting forces and the moments of inertia. In the present simulation the same agent mass for the entire population was chosen and a homogeneous mass distribution inside of each agent was assumed. Consequently, the moments of inertia differed from agent to agent according to their relative length.

Real bacteria live in a world of a very low Reynolds number[17]. Therefore, one may estimate that the coasting distance of a swimming bacterium is about 0.1 Å[17]. Taking this into consideration, strong damping forces $F_v$ were employed, linear to the velocity of the agents but pointing in the opposite direction, $F_v = \mu \cdot m \cdot v$, with a damping constant of $\mu=0.001$/tick. The time resolution for the simulation was 100 ticks per second. Due to this, an agent which moves with a velocity of 20 $\mu$m/s moves 20 Å per tick (about 4% of the average K-12 length[14]). In a subsequent tick a coasting agent can move only 0.02 Å.

The mass, $m$, of each agent was constant during the simulation and a known quantity from the beginning of the simulation and, as mentioned above, it was chosen the same for all agents. Using this mass and assuming a homogeneous mass distribution inside the ellipse-shaped, 2D planar polygon representing the bacterium, the value of the moment of inertia $I$ was calculated using Green's Theorem.

Aside from the damping forces, it is known that diffusion plays an important role in the world of real bacteria[18]. In particular one can observe that bacteria, that should actually move straight, meander as a consequence of rotational diffusion[18]. For that reason, the mean square angular deviation was approximated at the time $t = 1s$ according to the literature value for *E. coli* K-12, $<\alpha^2> = 2 D_R t = 27°$, where the rotational diffusion constant is $D_R = 0.124$ rad$^2$/s [18]. In the case of linear diffusion $D_L = 2 \cdot 10^{-13}$ cm$^2$/s was employed, also borrow from the literature of a paralyzed *E. coli* K-12 mutants[18].

Shape and momentum of inertia for each Bacterium was represented by a state vector containing information about the position and orientation of the agent. A second vector, i.e., the derivative of first state vector, contained information about the linear velocity and angular speed for the agent. Finally a third vector containing the second derivative of the first state vector was employed, accounting for the linear- and the angular acceleration. These three vectors were updated for every tick of the simulation corresponding to an update rate of 100 updates per second. The forces and torques that influenced all these properties are the following:

1. A driving force, $F_f$, which simulated the thrust of the flagella. This force is proportional to $F_f \approx v^\beta$, where $v$ is the velocity of the agent and $\beta$ is an empirical constant which is adjusted to reproduce the experimental data.
2. A collision force, $F_c$, which originates from inelastic collision between the agents. This force was modelled following the rules of classical mechanics for ridged bodies[19].
3. Damping, $F_v = \mu \cdot m \cdot v$, which simulate the dominant viscous forces of the laminar flow with the damping parameter $\mu$ the agent mass $m$ and the agent velocity $v$.
4. A forces corresponding to linear diffusion, $F_D$. The magnitude of $F_D$ was chosen to move a non-motile agent about a squared distance of $<x^2> = 2\,D_L\,t$,

Additionally to these forces, the agents experience moments of force (torques) which are:

1. A torque $T_f$, which simulates the flagella generated torque during the tumble movement. $T_f$ was chosen to reproduce the experimental turn angle distribution of swimming *E. coli*.
2. A torque $T_c$, which was produced by the collision of agents. As for the collision forces, $F_c$, the collision torques were modelled following the rules of classical mechanics for ridged bodies[19].
3. Damping, $T_v \approx \mu \cdot I \cdot \omega$, which simulates the viscous forces acting on the rotational movement of the agent with the damping parameter $\mu$ the moment of inertia $I$ and the angular speed $\omega$.
4. A torque corresponding to rotational diffusion $T_R$, where $T_R$ was chosen to turn a non-motile swimmer agent of about 1.5 $\mu$m length about a squared angle of $<\alpha^2> = 2\,D_R\,t$.

After summing up all force and torques that act during a tick on an agent, its position, as well as its velocity, was integrated using the standard Euler method. If in this context a collision was detected, the Euler integration was stopped, and the collision position was used as the initial condition to restart the Euler method. In this manner, a final position and a final velocity were found (in max 5 iterative steps), and both, position and velocity, were therefore updated.

In order to have a fast feedback of the simulation an openGL interface was programmed and used in order to show the movements of the bacterial colony and some observables, like the mean speed, turn angle distribution, etc., as shown in Figure 1. In addition all relevant data of the simulations were stored in output files. The final analyses of the simulations were performed on the basis of these output files using Wolfram Mathematica®.

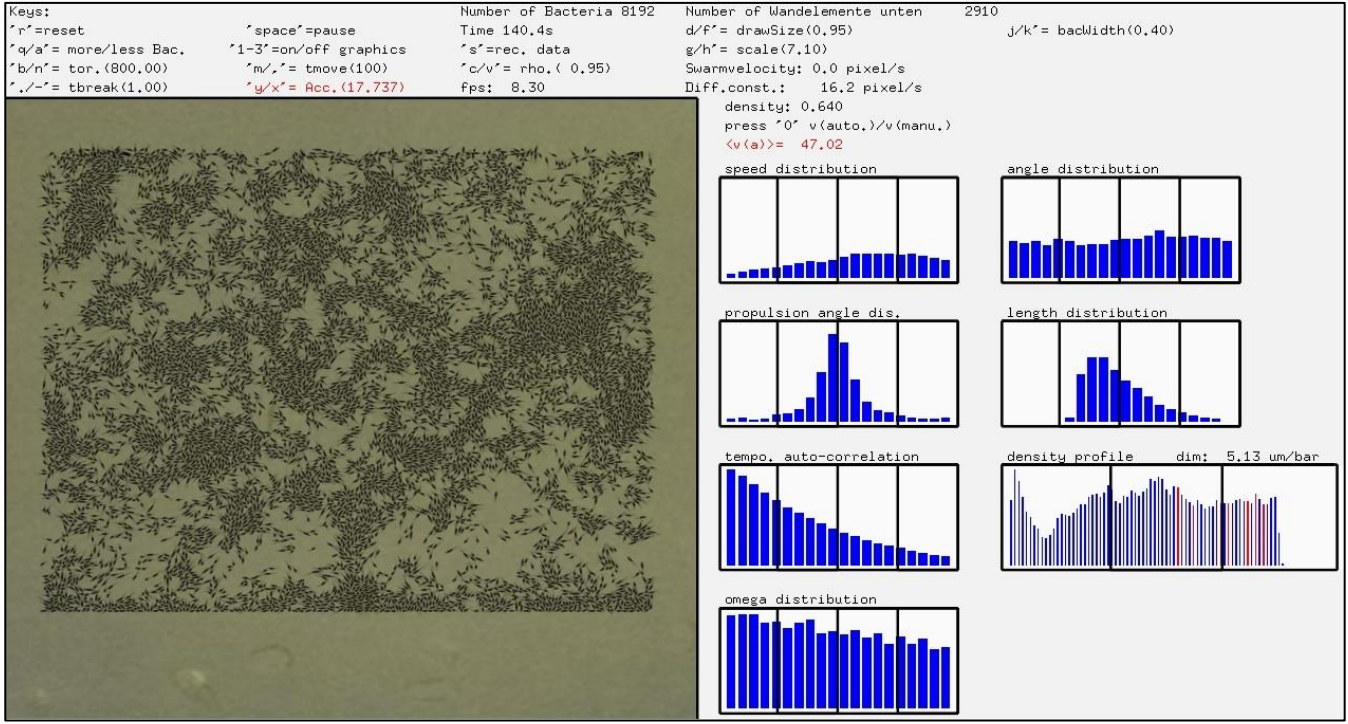

Figure 1. The image shows a screen shot of the simulation interface. At the left side the interface shows the swarming bacteria and at the right side, statistical observables used for the analysis of the simulation are displayed.

## III. RESULTS AND DISCUSSION

The hypothesis of this work is that swarming bacteria may exhibit run and tumble movements, but that these movements are completely superimposed by collisions among bacteria within the colony. It was therefore essential to simulate satisfactorily the bacterial swimming dynamics first. The run and tumble movement of individual bacteria was simulated using self-propelled agents which were moving freely without any interaction with other bacteria. These agents interacted with nothing but their environment via damping and diffusion. To ensure the quality of this basics dynamics, the results were compared to experimental data measure by Berg and Brown[3]. Table 1 shows the comparison between experimental results and simulations for some selected observables. In general, the simulation is in good agreement with these experimental results, but for the purposes of the present work a comparison of the obtained data, not only with single values, but also to different distributions was carried out.

Table 1 Experimental observables measured for real swimming *E. coli* K-12[3] and their corresponding simulated values (simulated agents).

|  | Real *E. coli* | Simulated agents |
|---|---|---|
| Mean speed [$\mu m$/s] | $14.2 \pm 3.4$ | $15.1 \pm 3.6$ |
| Mean tumble length [s] | $0.14 \pm 0.19$ | $0.15 \pm 0.01$ |
| Mean run length [s] | $0.86 \pm 1.18$ | $0.84 \pm 0.03$ |
| Mean direction change[°] | $68 \pm 36$ | $71 \pm 53$ |
| Diffusion constant [$\mu m^2$/s] | $100 \pm 106$ | $97 \pm 9$ |

On the basis of the results published by Berg and Browns[3] an experimental speed distribution of swimming *E. coli* was calculated. The speed distribution was obtained tracking the speed of a single bacterium for about 30 seconds. Due to the reduced set of data, the statistical significance of this experiment is limited, but it reveals that a bacterium may move during its run with relatively narrow distributed speed. In other words, the bacterium may reach an almost constant speed rather fast after its tumble and slow down also rather quickly before its next tumble. In the present simulations this behaviour is quasi automatically achieved as a consequence of the balance between damping forces $F_v$ and forward pushing flagella forces $F_f$. Although the average speed of the tracked bacterium is higher than the mean speed given in Table 1, an agent in the ensemble of "swimmers" was also found, thus showing a very similar speed distribution, as depicted in Figure 2.

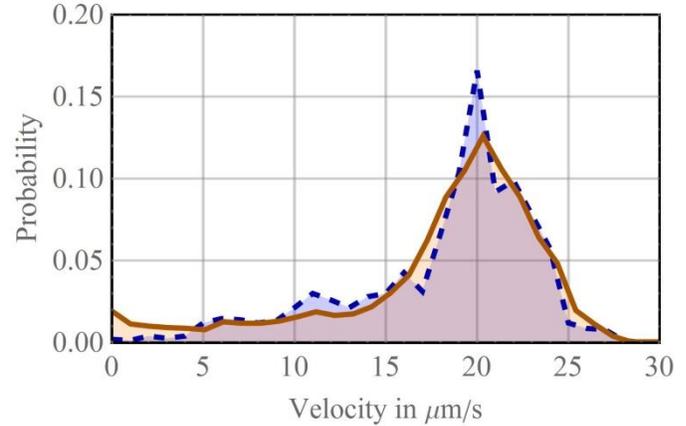

Figure 2 Experimental velocity distribution of a swimming *E. coli* bacterium (blue, dashed) calculated from the experimental data of Berg and Brown[3]. Corresponding simulated velocity distribution of a swimming agent (orange, solid).

The tumble length and the run length that Berg and Brown measured are exponentially distributed[3]. Using the experimental mean tumble length and mean run length give in Table 1, both exponential distributions were reconstructed and used for the present work to sample the run length and the tumble length for the agents. Consequently both, the run and the tumble lengths of the simulation were found to be in good agreement with the experimental results (not shown). The experimental expectation value of *E. coli* K-12s tumble angle, measured by Berg and Brown, was found to be about $68°$[3]. In addition to this expected value Berg and Brown published the corresponding tumble-angle distribution[3]. In contrast to simulations which employed circular shaped agents[20], the present work used rod shaped agents. This difference made it impossible to sample the tumble-angle of the agents directly from the experimental distribution, since an instant direction change would have led to unwanted overlaps among agents once the population density was increased (to simulate swarming). Instead, torques were applied on the agents in such manner that the resulting tumble angles of the simulation reproduced the experimental tumble angle distribution. This was challenging, since the tumble-angle distribution depends on the applied torque, the damping forces, the size of the bacteria (rotational inertia) and the tumble length. A comparison of the simulated and the experimental tumble-angle distribution is shown on the right side of Figure 3.

Since the here performed simulations were in good agreement with the experimental results for both, run and tumble movements, it was considered that the simulation reproduces the dynamics of swimming *E. coli* in low population densities satisfactorily. Based on this run and tumble movements bacterial swarming was simulated rising the population density.

The results of the swarming simulation were also analysed comparing observables of the simulation to experimental results. In this sense, it is important to take into account, that the swimming simulation was optimized to reproduce Bergs and Browns measurements on *E. coli* K-12. The work of Darton[14] provides a useful and comprehensive set of experimentally measured observables using swarming *E. coli* K-12, that were taken into account for comparison to the results of the present work. The setup used for swarming simulation differs only in three points from the setup of the swimming simulation.

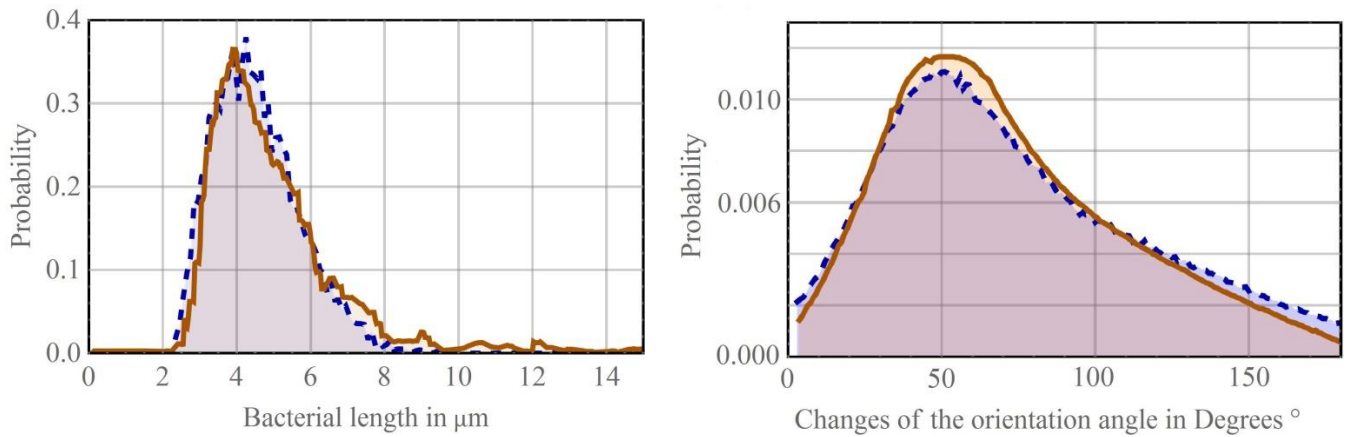

**Figure 3** Left: Experimental length distribution of the swarming *E. coli* K-12 (blue, dashed) and the length distribution used for the simulation (orange, solid) of the plateau with a population density of $\rho$=0.3. Right: Experimental tumble angle distribution of swimming *E. coli* K-12[3] (blue, dashed) and the tumble angle distribution obtained from the simulation (orange, solid).

The first point is the size distribution of the bacteria. As already mentioned real *E. coli* K-12 is a rod-shaped bacterium that is about 1 μm in diameter by 2 μm long when grown in sufficiently porous agar[14]. But when *E. coli* are inoculated on a moderately soft agar medium the bacteria elongate (and generate more flagella) and eventually start swarming[14]. The elongation from 2 μm to about 5.2 μm (the bacterial diameter is constant) changes the rotational inertia of the bacterium and therefore should influences the turn angle distribution. Such changes were considered in the swarming simulation using the experimental size distribution of the bacteria published by Darton[14] which are shown at the left side of Figure 3.

The second point concerns the quantity of bacteria per surface (population density), a rather obvious difference between the swimming setup and the swarming setup. Darton[14] distinguishes different zones within the swarming colony with different bacterial densities. The zone with the lowest density, called plateau, is in the centre of the colony and has a density of about 1/3. The density rises to a maximum of about 2/3 in a zone called peak close to the outer limit of the colony. The last zone, called edge, refers to the outer limit of the colony and shows a density between the densities of the plateau and the peak[14]. In order to compare the swarm model with Dartons results[14], different densities were used to simulate the three mentioned zones of the bacterial colony.

The third point addresses a change in the flagella force. This change was necessary to implement, since the collision among the agents leads to a slowdown of the average agent speed. This slowdown is usually not observed in experiments since bacteria may generate more force due to the generation of more flagella[10]. On the edge of the colony, where the monolayer of moving cells ends, the velocity distribution suggests a significantly different dynamics than in the inner of the colony. Therefore this peculiarity of the edge was simulated using very high friction for agents that move over a previously specified boarder which marks the end of the colony.

Except for these three changes, the entire swarming dynamics emerges from simple run and tumble swimming dynamics and inelastic collisions among agents. Figure 4 and Figure 5 show a comparison between the experimentally measured velocity distributions of Darton[14] and the velocity distributions generated by the present model for the different zones within the colony. One may notice that the expectation value of the velocity rises from the plateau, about 40 μm/s (shown on the right side of Figure 4), to a maximum at the peak, about 50 μm/s (Figure 5Figure 4), and falls from the peak to a minimum at the edge (shown on the left side of Figure 4). The modelled velocities and the experimental measurements are in good agreement, except for the velocities measured at the peak, where the simulation overestimates the number of fast bacteria. Possibly this is because the thin layer of liquid where bacteria move in is significantly thicker at the peak. The increasing degree of freedom may allow bacteria to go over each other and thus reduces collision, an effect that the present 2D model did not consider. However, using a simulation setup, where modified agents generate a lower driving force $F_f$, and less torque $T_f$, but have as well lower masses and moments of inertia, it was possible to archive a good agreement between experimental and simulated velocity distributions (shown on the right side of Figure 5). In this case the velocity distributions were also in good approximation with the Rayleigh distribution (using a mode value of about $\sigma \approx 41$). This indicated that the bacterial velocity within the peak zone of the colony may be described by a 2-dimensional vector of 2 orthogonal components which are uncorrelated, normally distributed with equal variance, and zero mean. Unfortunately simulations with these modified agents were not in agreement with other observables like the velocity-velocity autocorrelation, and for this reason they were not employed to perform further simulations.

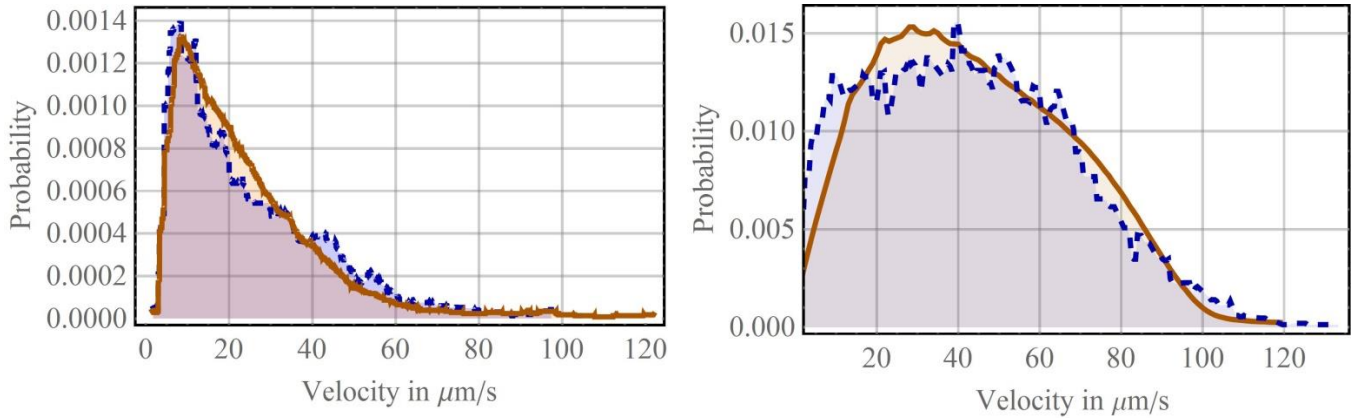

**Figure 4 Left:** Comparison between the experimentally measured velocity distributions of Darton[14] (blue, dashed) and the velocity distributions generated by our simulation (orange, solid) for the edge zone of the colony **Right:** Comparison between the experimental velocity distributions (blue, dashed) and the simulation (orange, solid) for the plateau with a population density of $\rho$=0.3.

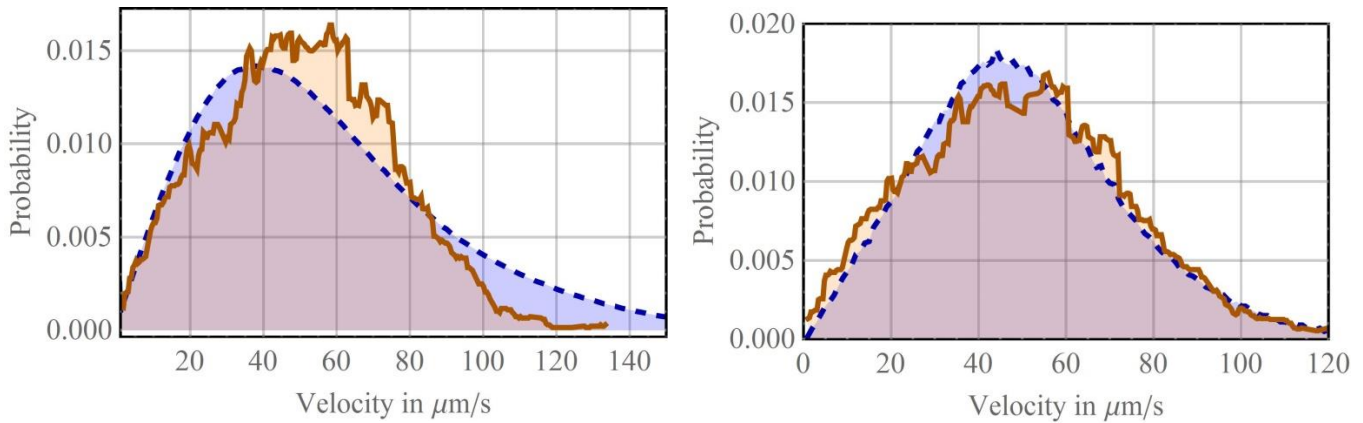

**Figure 5 Left:** The experimentally measured velocity distributions (orange, solid) and the velocity distributions generated by our simulation (blue, dashed) for the peak zone $\rho$=0.66. **Right:** The experimentally measured velocity distributions (orange, solid) and the velocity distributions generated by a simulation with modified parameters; bacteria have a lower mass, a lower moment of inertia and a lower flagella force. While the velocity distribution of the modified simulation is in very good agreement with the experimental data, the velocity-velocity autocorrelation decayed much slower than expected.

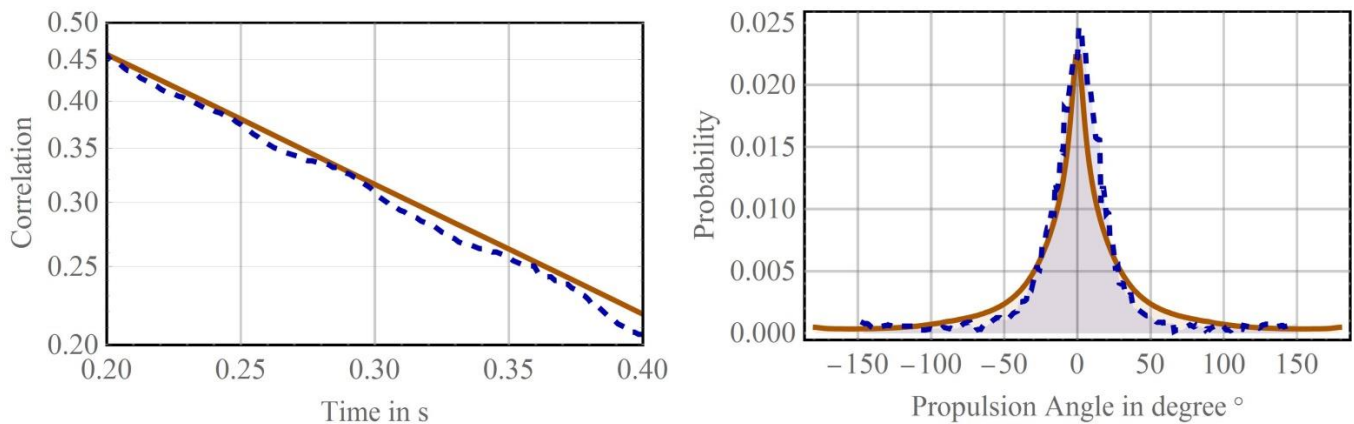

**Figure 6** Left: Velocity-velocity temporal autocorrelation of swarming bacteria (dashed blue) experimental function, (solid, orange) simulation with a population density of $\rho$=0.3. Right: Simulated (solid, orange) and the experimentally measured angular propulsion (dashed, blue) for the plateau with a population density of $\rho$=0.3.

The simulation of the edge zone was as well in good agreement with the experimental measurements; both the simulation and the experimental data show an overabundance of slow agents close to the edge of the colony since agents in this zone are frequently stalled as it was observed in the simulation and as well in experiments with real bacteria[14].

The plot on the left side of Figure 6 shows a comparison of the simulated and the measured velocity-velocity temporal autocorrelation function of the swarming bacteria. This autocorrelation determines the time over which the velocity of an agent becomes randomized. In both cases, the experimental and the simulated, the autocorrelation decays exponentially with a decay constant of $\lambda = 4$ s$^{-1}$.

The plot on right side of Figure 6 shows a comparison between the simulated and the experimentally measured angular propulsion. The propulsion angle was defined as the angle between the velocity vector and the orientation vector (from the tail to the head) of an agent. Both the experimental and the simulated propulsion angle distribution show that the propulsion is rather small and that the bacteria can swarm in the direction parallel to their orientation.

It must be remarked that an interesting fact, out of all features measured in this work, is the rather excellent agreement found between the results of the simulations and those of experimental measurements, especially taking into account that the swarm model employs only very few biological assumptions and that it is mainly based on a rather simple ridged body mechanics.

Simulations of system with agents that did not tumble were also carried out. In this case the agents performed runs in the same manner as swimming bacteria, stopped after a run length given by the run length distribution, but did not turn during the tumble length. Astonishing, the results of this simulations were identical to the results shown above (swarming agents with an individual run and tumble dynamics), expect for simulations where the active tumble-turns and additionally the rotational diffusion were switched off. In this case, the bacteria turned only because of inelastic collisions within the swarm, which led after a certain time to an enhance alignment of the agents, as shown in Figure 7. Using the expectation value of the argent orientation, one may notice that this value could be interpreted as an order parameter of the system, which tends, similar to the Vicsek-Model[21], to 1 when the system is aligned, without rotational diffusion, e.g., without noise, and to 0 for a system with randomly orientated agents, with rotational diffusion, e.g., with noise.

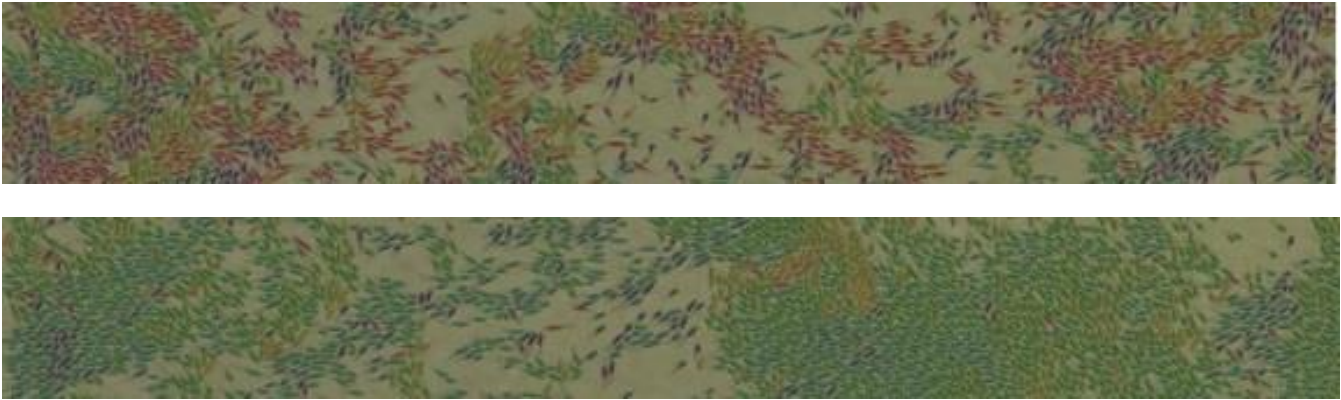

**Figure 7** The upper section of the figure shows a simulation of swarming bacteria, where individual agent is tuned to reproduce the individual movements (run and tumble) of swimming *E. coli*. It can be clearly seen, that the agents are orientated randomly (indicated by different colours). The lower section of the figure shows a simulation of swarming bacteria, where the individual agents perform modified run and tumble movements. Here active rotations of the bacteria during the tumble were turned off. Additionally rotational diffusion was turned off. It can be clearly noticed, that the bacteria are significantly more aligned.

The alignment was however favoured by the cyclic boundary conditions that were used for the simulations and which allowed the agents to move infinitely distances in one direction without reaching the boarder of the colony.

These results showed that the dynamics of swarming bacteria may very likely be dominated by the collisions of the agents, and that active turning agents may not be required (but rotational diffusion becomes mandatory) in order to reproduce the statistical observables shown above.

## IV. CONCLUSIONS

In this work we present a simple two-dimensional model which simulates the basic dynamics of swarming bacteria in a thin liquid film on the surface of a moderately soft agar plate. The model is based on self-propelled ridged bodies (agents), which interact with each other only through inelastic collisions and with the highly viscous environment thought damping forces and drag forces. The motions of individual agents are modelled closely corresponding to the behaviour of swimming bacteria. Accordingly, simulations with loosely packed agents ($\rho \approx 0$) show typical run-and-tumble statistics. In contrast, simulations with densely packed agents ($\rho \approx 0.3$-$0.7$) are dominated by interactions (collisions) between agents which lead to the emergence of a swarming behaviour. Simulations were also performed using *modified* agents that moved according to an altered swimming behaviour where active rotational movements of the agents were suppressed, but passive turns through rotational diffusion and collisions allowed. Both simulations show very good agreements with the literature values of real swarming Escherichia coli K-12[14]. In contrast, simulations employing densely packed *modified* agents in environments without rotational diffusion did not show the swarming behaviour of *E. coli*. Instead, these simulations showed an enhanced alignment of bacteria and the generation of large bacterial rafts which move all in one direction. This alignment might favoured by the cyclic boundary conditions that were used for the simulations and which allow the agents to move infinitely distances in one direction without reaching the boarder of the colony.

The present work clearly shows the importance of collisions in colonies of swarming *E. coli*, which determine dynamics of swarm. Furthermore, the present simulation results reveal, that the agents do not even have to rotate actively (passive rotational diffusion is mandatory) to reproduce the statistical observables measured by Darton et al.[14]. This finding support the observation of Darton that in swarming *E. coli* K-12 colonies "*collisions with adjacent cells, rather than active reorientation by flagellar reversal, are the dominant way that cells change direction while swarming*"[14]. In addition the results here reported support the statement of Mariconda et al. that bacteria which are unable to reverse their flagellar motor and hence cannot move via run and tumble can show swarming "*if the surface is sufficiently wet, exclusively clockwise or counterclockwise directions of motor rotation will support swarming, suggesting also that the bacteria can move on a surface with flagellar bundles of either handedness.*"[22]

## ACKNOWLEDGEMENTS

The authors acknowledge Consejo Nacional de Investigaciones Científicas y Técnicas (CONICET), Argentina, PIP 2014/16 N 112-201301-00629, for the financial support. DH and RCB are members of the research staff of CONICET. GF is grateful to CONICET for his postgraduate scholarship.